\documentclass[letter,
               keeplastbox,   
               ]{jacow}
\usepackage{pdfpages,multirow,ragged2e} %

\usepackage{lineno}

\usepackage {xcolor, afterpage}
\newcommand{\mm}{\mathrm}

\newcommand{\e}{\emph}
\usepackage[colorlinks=true,linkcolor=black,anchorcolor=black,citecolor=black,filecolor=black,menucolor=black,runcolor=black,urlcolor=black]{hyperref}
\makeatletter%
	\ifboolexpr{bool{xetex}}
	 {\renewcommand{\Gin@extensions}{.pdf,%
	                    .png,.jpg,.bmp,.pict,.tif,.psd,.mac,.sga,.tga,.gif,%
	                    .eps,.ps,%
	                    }}{}
\makeatother

\ifboolexpr{bool{xetex} or bool{luatex}} 
 {}                                      
 {\usepackage[utf8]{inputenc}}           

\usepackage[USenglish]{babel}

\ifboolexpr{bool{jacowbiblatex}}%
 {%
  \addbibresource{jacow-test.bib}
  \addbibresource{biblatex-examples.bib}
 }{}
\begin{document}

\title{The LCLS-II Gun \& Buncher LLRF Controller Upgrade}

\author{C. Bakalis\thanks{Main/Corresponding Author: \href{mailto:cbakalis@lbl.gov}{cbakalis@lbl.gov}}, L. Doolittle, D. Filippetto, G. Huang, S. Paiagua, C. Serrano\textsuperscript{1} \\
		A. Benwell, D. Chabot, J. Dusatko\textsuperscript{2} \\
    \textsuperscript{1}Lawrence Berkeley National Laboratory, Berkeley, CA 94720, USA \\
		\textsuperscript{2}SLAC National Accelerator Laboratory, Menlo Park, CA 94025, USA}
	
\maketitle

\begin{abstract}

   LCLS-II is currently in its commissioning phase at SLAC. It is an X-ray FEL driven by a CW superconducting LINAC. The beam injector plays a crucial role in the overall performance of the accelerator, and is critical to the final electron beam performance parameters. The LCLS-II injector comprises of a $185.7 \, \mm{MHz}$ VHF copper gun cavity, and a $1.3 \, \mm{GHz}$ two-cell L-band copper buncher cavity. The FPGA-based controller employs feedback and Self-Excited Loop logic in order to regulate the cavity fields. It also features several other functionalities, such as live detune computation, active frequency tracking, and waveform recording. The LLRF system drives the cavities via two $60 \, \mm{kW}$ SSAs through two power couplers, and thus stabilizes the fields inside the plant. This paper provides an outline of the general functionalities of the system, alongside a description of its hardware, firmware and software architecture, before finalizing with the current status of the project and its future goals.
\end{abstract}

\section{Introduction}
  LCLS-II is a Free-Electron Laser (FEL), operating at $4 \, \mm{GeV}$ via a Continuous-Wave (CW) superconducting LINAC, designed to provide pulses of electrons of $300 \, \mm{\upmu A}$ at up to $\sim 1 \, \mm{MHz}$. The soft X-rays delivered by the beam and its undulators are expected to fall within a spectral range of $200 - 5000 \, \mm{eV}$, thus allowing the newly-built machine to host a broad set of groundbreaking experiments \cite{lcls2}. The electron gun deployed at LCLS-II is a descendant of the Advanced Photo-injector EXperiment (APEX), which was a Very-High-Frequency gun (VHF-Gun), capable of CW operation and optimized for the performance required by MHz-rate X-ray FELs \cite{apex}. Alongside the LCLS-II injector system, the High Repetition-rate Electron Scattering (HiRES) experiment at LBNL shares the same gun and buncher concepts \cite{hires} - this sets a fundamental requirement for the controller under review in this paper, which is to be able to serve both of these experiments without major modifications in any of its hardware, firmware, or software components.

  The injector of LCLS-II uses APEX-style RF cavities, including a $185.7 \, \mm{MHz}$ VHF copper gun cavity and a $1.3 \, \mm{GHz}$ two-cell L-band copper buncher cavity, with a loaded Q in the order of $15000$ \cite{gang}. HiRES on the other hand uses a similar scheme, with the critical frequencies of all aforementioned setups summarized in Table~\ref{tab:freq}:

  \begin{table}
    \centering
    \caption{Several important frequencies of all controller configurations}
    \begin{tabular}{|c||c|c|c|}
    \hline
    \hline
    $\mm{MHz}$ & LCLS-II Gun & LCLS-II Buncher & HiRES Gun \\
    \hline
    \hline
    $f_{LO}$ & $1320$ & $1320$ & $1225.7$ \\
    \hline
    $f_{ADC}$ & $94.286$ & $94.286$ & $102.14$ \\
    \hline
    $f_{DAC}$ & $188.57$ & $188.57$ & $204.29$ \\
    \hline
    $f_{RF}$ & $185.7$ & $1300$ & $185.7$ \\
    \hline
    $f^{up}_{IF}$ & $34.285$ & $145$ & $18.571$ \\
    \hline
    $f^{dn}_{IF}$ & $20.714$ & $20$ & $74.288$ \\
    \hline
    $f_{PRL}$ & $1300/7$ & $1300$ & N/A \\
    \hline
    \hline
    \end{tabular}
    \label{tab:freq}
    \end{table}

\section{Controller System}
To ease the development process, all components of the LCLS-II injector controller share the same hardware, firmware and software base as the SRF part of the system. The chassis for the gun control is different than the one used for the buncher. The basic function of the LLRF system is to stabilize the RF field in the cavities with feedback, and allow the operator to monitor the RF signals by via a waveform recording and display scheme. In the LCLS-II case, a high precision phase reference is distributed globally by a hard coaxial cable. The phase of the PRL is digitally averaged to provide a drift-compensated phase reference. The system needs to fulfill stringent requirements for the stability of the cavity fields. The gun cavity needs to be within a $0.01 \%$ field amplitude error, and a $0.015^\circ$ phase error. The buncher cavity needs to be within a $0.3 \%$ field amplitude error, and a $0.05^\circ$ phase error \cite{gang}. What follows is a brief description of each system component.

\subsection{Hardware}
\begin{figure}[!htb]
   \centering
   \includegraphics*[width=.5\columnwidth, angle=180]{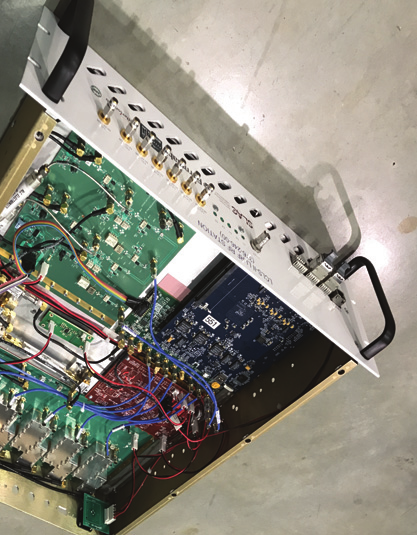}
   \caption{Gun/Buncher Chassis Assembly. The FPGA motherboard is the blue board on the bottom-left, while the digitizer can be seen directly attached to it on its bottom. The analog mixers are located on the far back of the chassis, next to its back-plane.}
   \label{fig:chassis}
\end{figure}

The foundation of the hardware is based on a chassis, comprised of devices operating both in the analog and in the digital domain. The chassis can be seen in Figure~\ref{fig:chassis}. The LO of the system is driven to one of the chassis inputs, and is then distributed to various parts of the chassis. It is used in analog downmixers, utilized in all of the chassis inputs, which include cavity signals or the PRL. The analog mixers provide downconverted signals to the digitizer board, which also accepts the LO as a clock reference. The digitizer employs a clocking chip that uses the LO as an input, and in turn distributes a reference relevant to the setup at hand, depending on its configuration, which is software-controlled. It also deploys ADCs to sample the downmixed RF signals, and DACs that receive a digital stream before converting them to an IF voltage. Naturally, the analog part of the chassis also has upmixers that receive the said DAC stream before converting it to RF that will drive the SSAs outside of the chassis. The digitizer is the daughter (or mezzanine) board of the purely digital board, which houses the controller firmware in its FPGA. The FPGA receives the digitized ADC samples, and uses them to control the cavity fields via its DAC output. The FPGA is also connected to the Ethernet interface, in order to connect to the software.

\subsection{Firmware}
\begin{figure}[!htb]
   \centering
   \includegraphics*[width=\columnwidth]{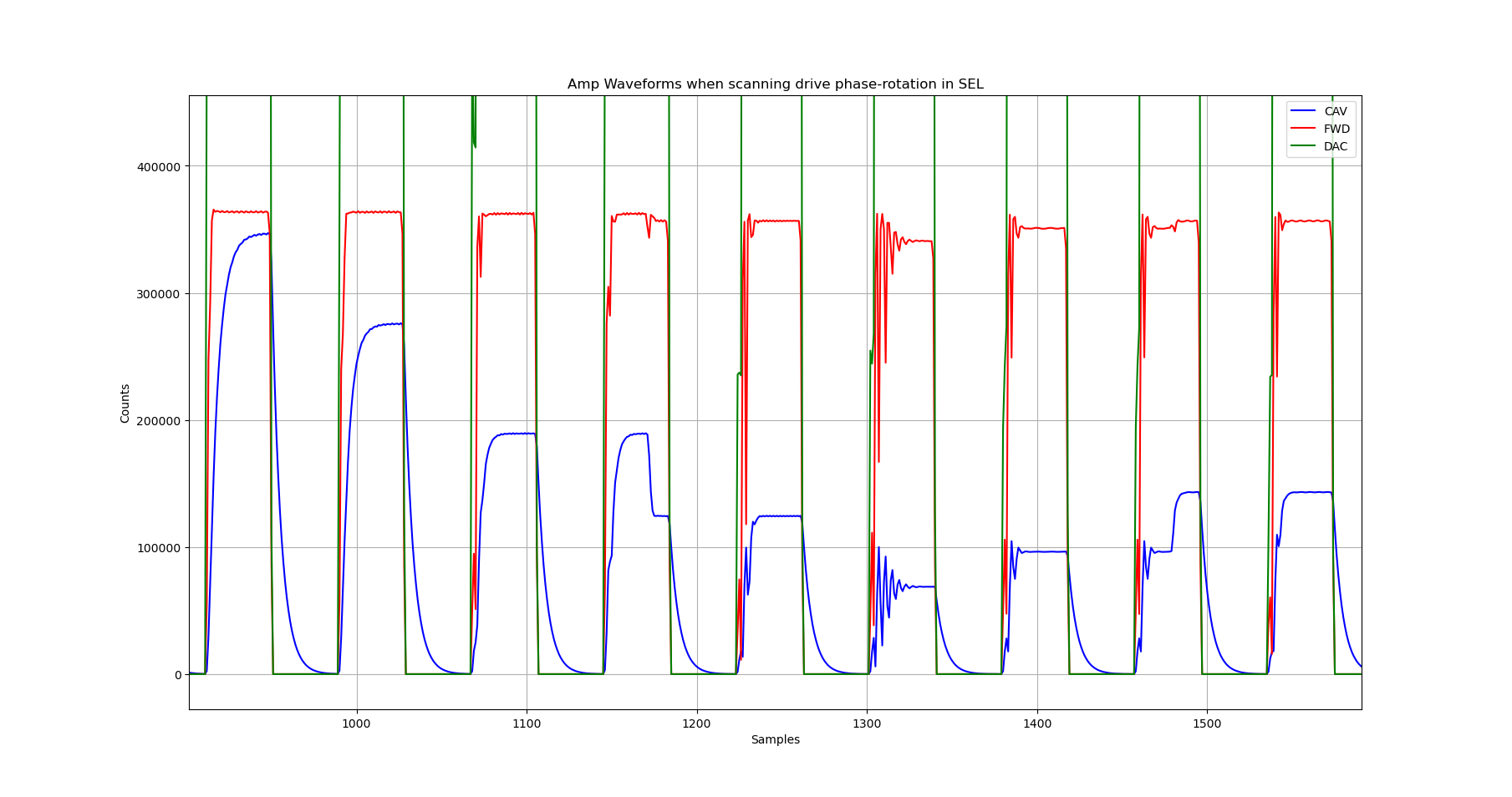}
   \caption{SEL phase scan when running using the Verilog-based cavity emulator. Different pulses are associated with different phase offsets, while the cavity probe and forward signals that are clearly visible here are generated by the emulator.}
   \label{fig:selscan}
\end{figure}

\begin{figure*}
      \centering
      \includegraphics*[width=\textwidth]{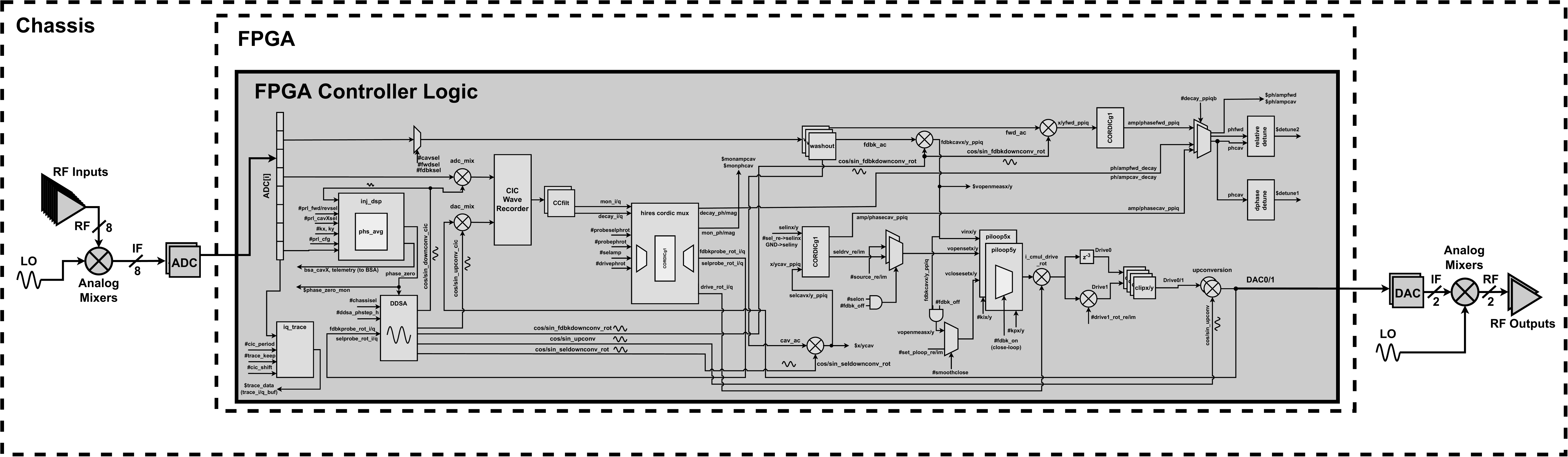}
      \caption{Block diagram of the LCLS-II Gun/Buncher controller firmware.}
      \label{fig:fw}
  \end{figure*}

A block diagram of the firmware design can be seen in Figure~\ref{fig:fw}. The firmware receives the ADC samples via a serial stream, which is deserialized using the same clock as the one used by the ADCs as a reference. This defines the one main domain of the design, alongside the DAC clock domain, which is shared with the DACs on the digitizer. The $16-$bit ADC samples are used in various parts of the controller firmware. Digital mixers/multipliers bring the signals down to the baseband level, and drive them to a CIC filter which provides a stream of waveform data to the user via the Ethernet interface and the software infrastructure. The capturing is triggered depending on the configuration of the firmware, and is usually aligned with the falling-edge of the outbound RF pulse when operating in non-CW mode. The mixers that drive the CIC filter, and all of the mixers in the design in general, are driven by digitally-generated harmonics, provided by a local phase-accumulator coupled with a CORDIC module. The phase-accumulators determine the frequencies of the signals that drive the mixers, and is very important for their parameters to be versatile, since the design needs to support several downconverting and upconverting frequencies (see Table~\ref{tab:freq}). There are also several demultiplexers that allow for selecting different ADC streams, and consequently drive them to sub-components that manage different operations. These selection parameters are also software-controllable, and vary depending on the setup. The core of the controller lies in its Self-Excited Loop (SEL) and its feedback logic. There are two different inputs to these parts, and the user can select which cavity probe to use for each. Usually, the same is used for both, but an independent configurable phase offset is applied to the feedback cavity input. During the cavity bring-up procedure, the firmware scans for the optimal SEL phase offset, and uses that when running in SEL mode. To ease the development process, a Verilog-based (non-synthesizable) cavity emulator was deployed, interfacing with the controller's cavity field and drive when running HDL simulations. This facilitates the debugging procedure greatly, and in Figure~\ref{fig:selscan} one can see the results of the SEL phase scan using the firmware-based emulator alone.
\vspace{0.2cm}

During operation, the firmware engages its feedback loop in order to stabilize the cavity field. Finally, the firmware can also calculate the relative detune of the cavity, which is published to the software host, which in turn uses that to fine-tune the phase-accumulators of the firmware, to optimize performance, using a simple slow-control loop. This is especially important for the copper gun cavity, which changes substantially in terms of its resonant frequency during its warm-up process.

\subsection{Software}
\begin{figure}[!htb]
   \centering
   \includegraphics*[width=\columnwidth]{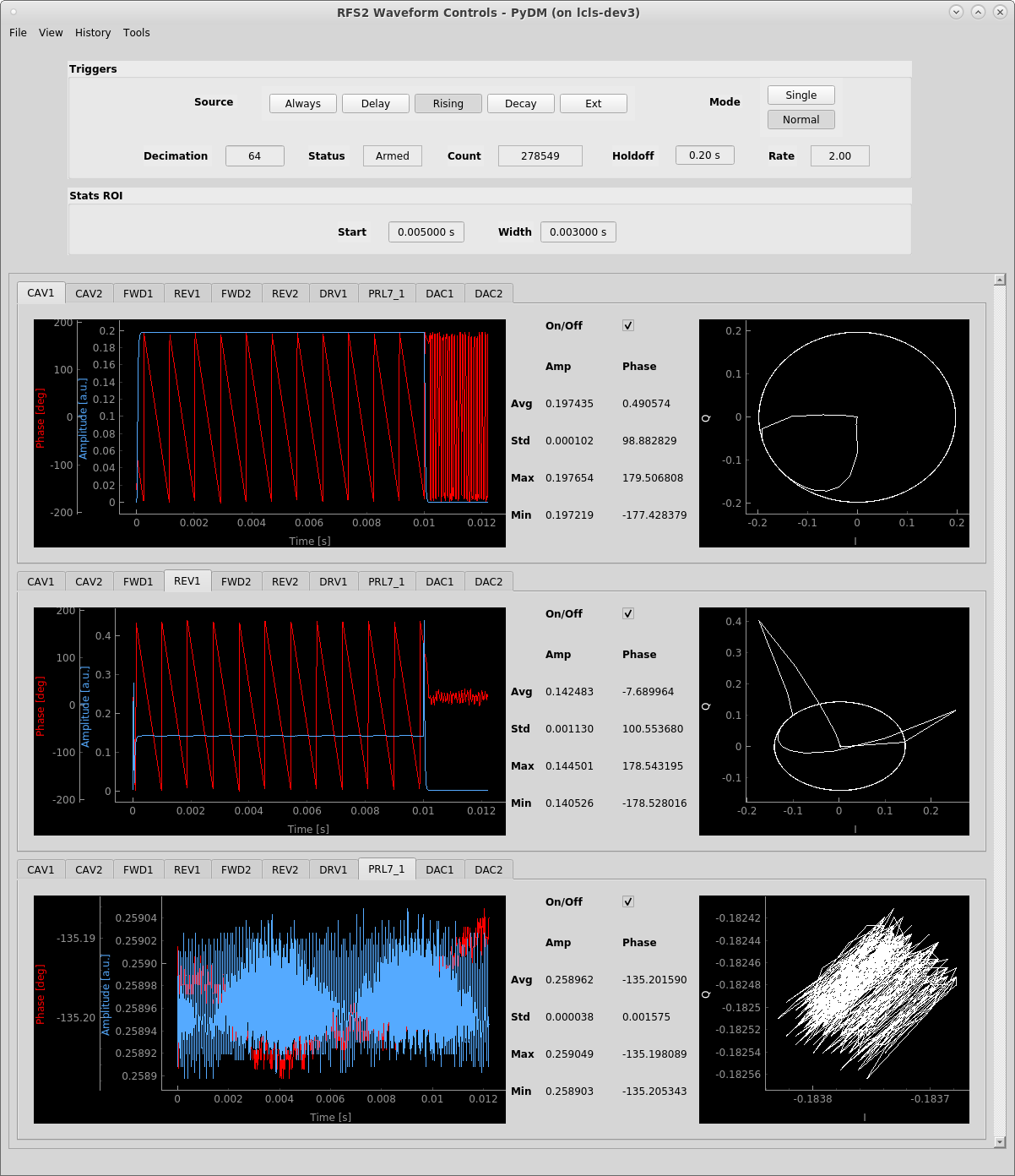}
   \caption{EPICS GUI during SEL operation.}
   \label{fig:epicsgui}
\end{figure}

High level control and integration is via the EPICS \cite{epics} software stack, supporting UDP/IP communication with board firmware as well as a general
user interfacing requirements. Graphical interfaces are available in both the Phoebus \cite{phoebus} and PyDM \cite{pydm} toolkits, reflecting the differing local standards between LBNL and SLAC. Both GUIs utilize identical lower-level support software, and provide the means for firmware (re-)initialization, full control of RF functionality and modes, as well as RF waveform visualization and statistics (see Figure~\ref{fig:epicsgui}). Independent of the EPICS tools, a suite of
Python software is also available for board programming, diagnostics, and calibration.

\section{Current Status and Conclusions}
Currently, the design is able to control the gun cavity at HiRES, and can also support hardware-based cavity emulators that are within the frequency and bandwidth ranges of the LCLS-II gun and buncher. It is foreseen to deploy the system on the LCLS-II setup in the coming months. What is also in the works is the separation of chassis, based on whether they are involved in the downconversion or upconversion of the RF signals, in order to enhance channel separation and thus improve performance.

\section{Acknowledgments}
All authors acknowledge support from the Office of Science, Office of Basic Energy Sciences, of the U.S. Department of Energy, under Contract No. DE-AC02-05CH1123

\ifboolexpr{bool{jacowbiblatex}}%
	{\printbibliography}%
	{%
	
	
} 

%
%


\end{document}